\documentclass[conference]{IEEEtran}

\usepackage{graphicx}
\usepackage{enumerate}
\usepackage{cite}
\usepackage{bm}
\usepackage[dvipsnames]{xcolor}
\usepackage{amsmath,amssymb,amsthm}
\usepackage{dutchcal}
\usepackage{stfloats}
\usepackage{suffix}
\usepackage{mathtools}
\newcounter{MYtempeqncnt}

\DeclareMathAlphabet{\mathpzc}{OT1}{pzc}{m}{it}

\renewcommand*{\arraystretch}{1.5}
\newcommand{\kron}{ \hspace{-.8mm} \otimes \hspace{-.8mm}}
\newcommand{\kr}{ \hspace{-.8mm} \odot \hspace{-.8mm}}

\usepackage[font=footnotesize,skip=2pt]{caption}
\DeclarePairedDelimiterX\MeijerM[3]{\lparen}{\rparen}%
{\begin{smallmatrix}#1 \\ #2\end{smallmatrix}\delimsize\vert\,#3}

\newcommand\MeijerG[8][]{%
  G^{\,#2,#3}_{#4,#5}\MeijerM[#1]{#6}{#7}{#8}}

\WithSuffix\newcommand\MeijerG*[7]{%
  G^{\,#1,#2}_{#3,#4}\MeijerM*{#5}{#6}{#7}}

\begin{document}

\title{IRS-Assisted Massive MIMO-NOMA Networks with Polarization Diversity \vspace{-3mm}}

\author{\IEEEauthorblockN{Arthur S. de Sena\IEEEauthorrefmark{1},
		Pedro H. J. Nardelli\IEEEauthorrefmark{1}, Daniel B. da Costa\IEEEauthorrefmark{2}, F. Rafael M. Lima\IEEEauthorrefmark{2},\\ Liang Yang\IEEEauthorrefmark{3}, Petar Popovski\IEEEauthorrefmark{4},
		Zhiguo Ding\IEEEauthorrefmark{5}, and Constantinos B. Papadias\IEEEauthorrefmark{6}}
	\IEEEauthorblockA{\IEEEauthorrefmark{1} Lappeenranta University of Technology, Finland \hspace{5mm}
	 \IEEEauthorrefmark{2} Federal University of Ceará, Brazil\hspace{5mm}
	 \IEEEauthorrefmark{3} Hunan University, China\\
	 \IEEEauthorrefmark{4} Aalborg University, Denmark \hspace{5mm}
	 \IEEEauthorrefmark{5} The University of Manchester, UK\hspace{5mm}
	 \IEEEauthorrefmark{6} The American College of Greece, Greece}
	{Emails: arthur.sena@lut.fi, pedro.nardelli@lut.fi, danielbcosta@ieee.org, rafaelm@gtel.ufc.br, } 
	\\{ liangy@hnu.edu.cn, petarp@es.aau.dk, zhiguo.ding@manchester.ac.uk, cpapadias@acg.edu \vspace{-5mm}}
}

\maketitle

\vspace{-20mm}
\begin{abstract}
In this paper, the appealing features of a dual-polarized intelligent reflecting surface (IRS) are exploited to improve the performance of dual-polarized massive multiple-input multiple-output (MIMO) with non-orthogonal multiple access (NOMA) under imperfect successive interference cancellation (SIC). By considering the downlink of a multi-cluster scenario, the IRSs assist the base station (BS) to multiplex subsets of users in the polarization domain. Our novel strategy alleviates the impact of imperfect SIC and enables users to exploit polarization diversity with near-zero inter-subset interference. Our results show that when the IRSs are large enough, the proposed scheme always outperforms conventional massive MIMO-NOMA and MIMO-OMA systems even if SIC error propagation is present. It is also confirmed that dual-polarized IRSs can make cross-polar transmissions beneficial to the users, allowing them to improve their performance through polarization diversity.
\end{abstract}


\IEEEpeerreviewmaketitle

\vspace{-2mm}
\section{Introduction}
	\vspace{-1mm}

Massive multiple-input multiple-output (MIMO) is one key technology for the fifth-generation (5G) wireless systems. The technology uses a large number of antennas at the base station (BS) to transmit parallel data streams to multiple users through spatially separated beams. Conventionally, orthogonal multiple access (OMA) techniques are combined with massive MIMO to guarantee zero inter-beam interference in scenarios where it is difficult to multiplex users in the space domain. Even though such schemes can effectively cope with the interference issue, they may perform poorly in terms of spectral efficiency and latency as the number of users increases. Therefore, MIMO-OMA systems are not ideal for ultra-dense deployments, and this motivates the use of non-orthogonal multiple access (NOMA), such that MIMO-NOMA can serve simultaneously several users with non-separable beams.

The performance of a massive MIMO-NOMA network scales up with the increase of antennas. However, due to physical space constraints, in practical systems the number of antennas is limited at both the BS and user's devices. One efficient strategy to alleviate such a limitation can be achieved by arranging the antenna elements into co-located pairs with orthogonal polarizations, forming a dual-polarized antenna array. With such an approach, it becomes possible to install twice the number of antennas of a single-polarized array utilizing the same physical space. In addition, dual-polarization enables massive MIMO-NOMA systems to exploit diversity in the polarization domain, which can significantly outperform conventional single-polarized schemes \cite{ni3}.

Despite the mentioned advantages, a dual-polarized massive MIMO-NOMA system still has limitations. For instance, the stochastic nature of the scatterer environment can depolarize the transmitted signals and generate cross-polar interference at the receivers. As demonstrated in \cite{ni3}, this depolarization phenomenon can deteriorate the system performance. Furthermore, in power-domain NOMA, the users need to employ successive interference cancellation (SIC) to decode their received data symbols, which also has some drawbacks. An increase in the number of users leads to higher interference and a more complex SIC decoding process, potentially resulting in excessive decoding errors, and lowered system throughput.

Therefore, new strategies and technologies are needed to overcome the above impairments. In this sense, the recent concept of an intelligent reflecting surface (IRS)~\cite{Renzo2020} holds a great potential. An IRS is an engineered device that comprises multiple sub-wavelength reflecting elements with reconfigurable electromagnetic properties. The phases and amplitudes of reflections induced by the IRS elements are controlled independently via software, which enables them to, collectively, forward the impinging waves with an optimized radiation pattern and reach diverse objectives like steering, collimation, absorption, and control of polarization \cite{aswc2020}. Such appealing features unlock countless new possibilities for manipulating the random phenomena of electromagnetic propagation, a critical issue in any wireless communication system. This is discussed in several recent works, some of them dealing specifically with MIMO-NOMA schemes \cite{rw01,rw02,rw05,rw06}.

Although IRSs have been studied in different scenarios and applications recently, to the best of our knowledge, all related works are limited to single-polarized systems, and there are no works that exploit IRSs for manipulating polarization in dual-polarized MIMO-NOMA networks. Motivated by this, in this paper, we harness the features of dual-polarized IRSs for enabling polarization diversity and for reducing the impact of imperfect SIC in a dual-polarized MIMO-NOMA network. The reflecting elements of each IRS are optimized to mitigate the transmissions originated at the BS from the interfering polarization. We transform the complicated original problem into quadratic constrained quadratic sub-problems, and we show that their optimal solutions can be obtained via interior-points methods in polynomial time. We present representative numerical simulation results alongside with comprehensive discussions. For instance, we show that when the IRSs are large enough, the proposed scheme always outperforms conventional massive MIMO-NOMA and MIMO-OMA systems even if SIC error propagation is present. 

{\bf Notation and Special Functions:} Bold-faced lower-case letters denote vectors and upper-case represent matrices. The $i$th element of a vector $\mathbf{a}$ is denoted by $[\mathbf{a}]_i$, the $(ij)$ entry of a matrix $\mathbf{A}$ by $[\mathbf{A}]_{ij}$, and the transpose and the Hermitian transpose of $\mathbf{A}$ are represented by $\mathbf{A}^T$ and $\mathbf{A}^H$, respectively. The symbol $\otimes$ represents the Kronecker product, $\odot$ is the Khatri-Rao product \cite{Brewer78}, $\mathbf{I}_M$ represents the identity matrix of dimension $M\times M$, and $\mathbf{0}_{M, N}$ denotes the $M\times N$ matrix with all zero entries. The operator $\textit{vec}\{\cdot \}$ transforms a matrix of dimension $M\times N$ into a column vector of length $MN$, the operator $\textit{vecd}\{\cdot \}$ converts the diagonal elements of an $M\times M$ square matrix into a column vector of length $M$, $\textit{diag}\{\cdot \}$ transforms a vector of length $M$ into an $M\times M$ diagonal matrix, and $\Re\{\cdot \}$ returns the real part of a complex number.
\vspace{-2mm}

\section{System Model}

Consider a single cell MIMO-NOMA network where one BS is communicating in downlink mode with multiple users. Both users and the BS comprise dual-polarized antenna elements that are arranged into multiple co-located pairs, each one containing one vertically and one horizontally polarized antenna element. The users are equipped with $N/2$ pairs of dual-polarized receive antennas, and the BS with $M/2$ pairs of dual-polarized transmit antennas, where we assume that $M$ and $N$ are even, and $M\gg N$. Moreover, the users are distributed among $K$ spatial clusters that are organized into $G$ groups of $U$ users each.

We program the BS to further subdivide each of the $G$ groups into two polarization subsets, namely vertical subset and horizontal subset, each one containing $U^{p}$ users, $p\in \{v,h\}$, i.e., $U^{v}$ users are served with vertically polarized transmit antennas, and $U^{h}$ users are served with horizontally polarized antennas, such that $U^{v} + U^{h} = U$. To enable this scheme, we optimize dual-polarized IRSs to ensure that signals transmitted from one polarization impinge only at users assigned to that specific polarization. For this, we assume that there are $U$ IRSs with $L$ dual-polarized reflecting elements installed within each group and that each IRS assists one user.

\begin{figure}
	\centering
	\includegraphics[width=.9\linewidth]{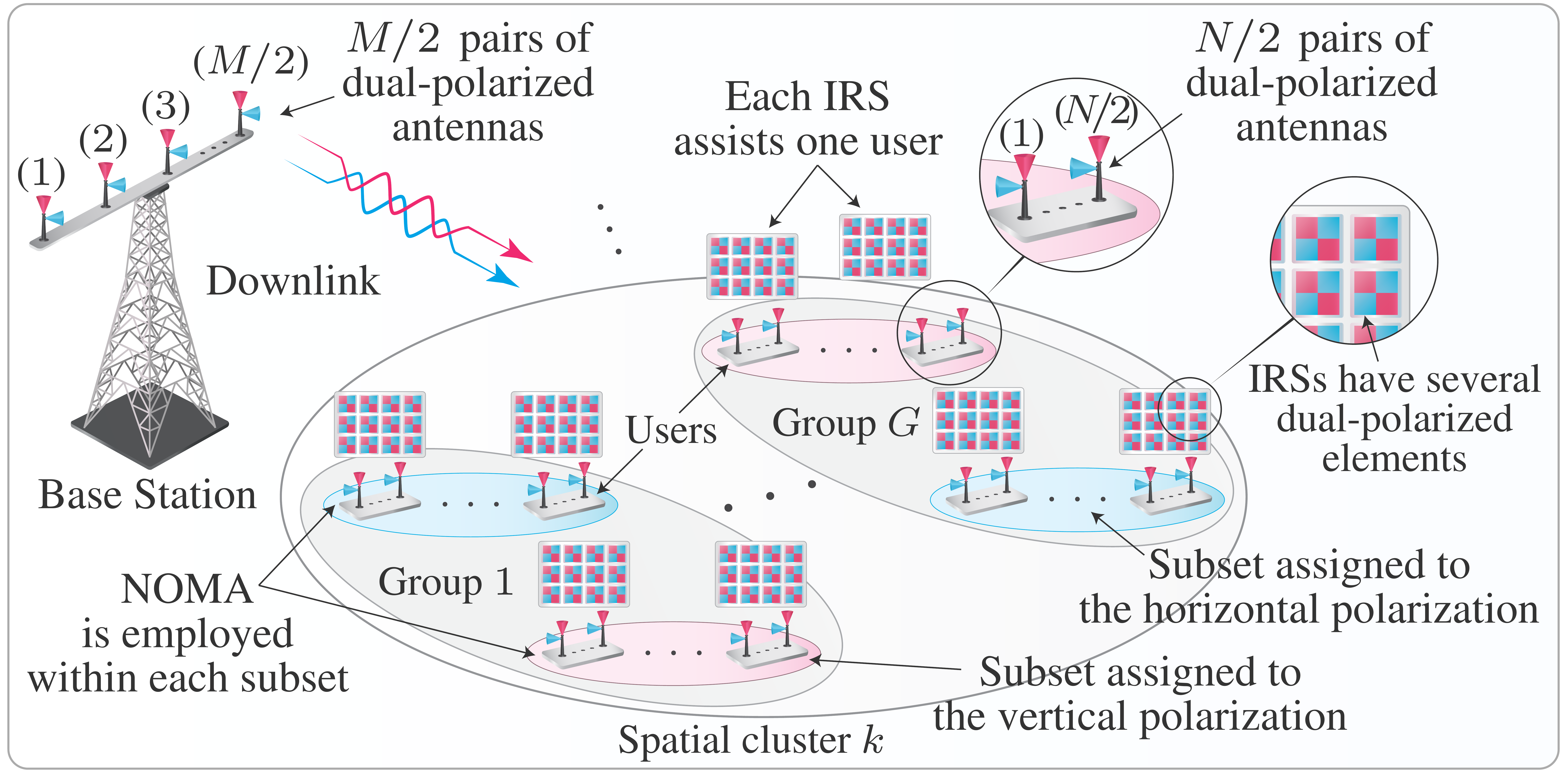}
	\caption{System model. Dual-polarized IRSs enable users to exploit polarization diversity by mitigating polarization interference.}\label{fig:sysmodel}
\end{figure}

Following the proposed strategy, the BS applies superposition coding to each polarization subset and transmit the superimposed messages through the assigned polarization. More specifically, the BS sends the following signal
\begin{align}\label{eq03}
\mathbf{x} \hspace{-.5mm}=  \hspace{-1mm} \sum_{k=1}^{K} \hspace{-.5mm} \mathbf{P}_k\hspace{-1mm}\renewcommand*{\arraystretch}{.8} \begin{bmatrix}
\mathbf{x}^v \\ \mathbf{x}^h
\end{bmatrix} \hspace{-.5mm} =\hspace{-.5mm} \sum_{k=1}^{K}\hspace{-.5mm} \mathbf{P}_k \hspace{-1mm} \sum_{g=1}^{G} \sum_{u=1}^{U} \hspace{-.5mm} \mathbf{v}_{kgu} 
\alpha_{kgu}{x}_{kgu} \hspace{-.5mm}\in\hspace{-.5mm} \mathbb{C}^{M \times 1},
\end{align}
where $\mathbf{x}^p$ is the data vector transmitted in the polarization $p \in \{v,h\}$. ${x}_{kgu} $ and  $\alpha_{kgu}$  are, respectively, the symbol and the power coefficient for the $u$th user in the $g$th group within the $k$th cluster. $\mathbf{P}_k \in \mathbb{C}^{M\times \bar{M}}$ is a precoding matrix intended to eliminate inter-cluster interference, in which $\bar{M}$ is a parameter that controls the number of effective data streams, and $\mathbf{v}_{kgu} \in \mathbb{C}^{\bar{M} \times 1}$ is a precoding vector responsible for multiplexing the users in the polarization domain, satisfying $\|\mathbf{v}_{kgu}\|^2 = 1$.

A dual-polarized IRS with $L$ elements can be modeled by a block matrix partitioned into four $L\times L$ diagonal sub-matrices, each one modeling reflections from one polarization to another. Specifically, the reflection matrix for the dual-polarized IRS that assists the $u$th user in the $g$th group of the $k$th cluster is:
\begin{align}\label{coeffmtx}
\bm{\Theta}_{kgu} = \renewcommand*{\arraystretch}{1}
    \begin{bmatrix}
        \bm{\Phi}_{kgu}^{vv} & \bm{\Phi}_{kgu}^{hv} \\
        \bm{\Phi}_{kgu}^{vh} & \bm{\Phi}_{kgu}^{hh}
    \end{bmatrix} \in \mathbb{C}^{2L\times 2L},
\end{align}
where {$\bm{\Phi}_{kgu}^{pq} = \textit{diag}\{[\omega^{pq}_{kgu,1} e^{-j\phi^{pq}_{kgu,1}},\omega^{pq}_{kgu,2} e^{-j\phi^{pq}_{kgu,2}}, \cdots,$ $\omega^{pq}_{kgu,L} e^{-j\phi^{pq}_{kgu,L}}]\} \in \mathbb{C}^{L\times L}$}, with $\phi^{pq}_{kgu,l}$ and $\omega^{pq}_{kgu,l}$ representing, respectively, the phase and amplitude of reflection induced by the $l$th element from polarization $p$ to polarization $q$, with $p,q \in \{v,h\}$, in which $|\omega^{pq}_{kgu,l}|^2 \leq 1$. By using the multi-polarized and the dyadic backscatter channel models \cite{ni3,Renzo2020}, the full channel matrix for the $u$th user in the $g$th group of the $k$th cluster can be structured as
\begin{align}\label{eq:ch0} 
    &\mathbf{H}^H_{kgu} = 
    \text{\footnotesize $\sqrt{\zeta_{kgu}^\text{\tiny BS-IRS} \zeta_{kgu}^\text{\tiny IRS-U}}
    \frac{1}{\sqrt{2}}\renewcommand*{\arraystretch}{1.4} \setlength{\arraycolsep}{1pt}
    \begin{bmatrix}
        \mathbf{\bar{S}}_{kgu}^{ vv} &
        \mathbf{0}_{L,\frac{N}{2}} \\
        \mathbf{0}_{L,\frac{N}{2}} &
        \mathbf{\bar{S}}_{kgu}^{ hh}
    \end{bmatrix}^H  \begin{bmatrix}
        \bm{\Phi}_{kgu}^{vv} & \bm{\Phi}_{kgu}^{hv} \\
        \bm{\Phi}_{kgu}^{vh} & \bm{\Phi}_{kgu}^{hh}
    \end{bmatrix} $} 
    \nonumber\\[-.5mm]
    &\text{\footnotesize $\times \hspace{-1mm} \renewcommand*{\arraystretch}{1.4} \setlength{\arraycolsep}{-3pt}
    \begin{bmatrix}
        \mathbf{\bar{G}}_{kgu}^{ v v} &
        \sqrt{\chi^{\text{\tiny BS-IRS}}}
        \mathbf{\bar{G}}_{kgu}^{ h v} \\
        \sqrt{\chi^{\text{\tiny BS-IRS}}}
        \mathbf{\bar{G}}_{kgu}^{ v h} &
        \mathbf{\bar{G}}_{kgu}^{ h h}
    \end{bmatrix} \hspace{-1mm} + \hspace{-.5mm}
    \sqrt{\zeta_{kgu}^\text{\tiny BS-U}}\hspace{-1mm}
    \begin{bmatrix}
        \mathbf{\bar{D}}_{kgu}^{ v v} &
        \sqrt{\chi^{\text{\tiny BS-U}}}
        \mathbf{\bar{D}}_{kgu}^{ v h} \\
        \sqrt{\chi^{\text{\tiny BS-U}}}
        \mathbf{\bar{D}}_{kgu}^{ h v} &
        \mathbf{\bar{D}}_{kgu}^{ h h}
    \end{bmatrix}^H\hspace{-1mm}$},
\end{align}
where $\mathbf{\bar{D}}_{kgu}^{ pq} \in \mathbb{C}^{\frac{M}{2} \times \frac{N}{2}}$, $\mathbf{\bar{S}}_{kgu}^{ pq} \in \mathbb{C}^{L\times \frac{N}{2}}$, and $\mathbf{\bar{G}}_{kgu}^{ pq} \in \mathbb{C}^{L\times \frac{M}{2}}$ model, respectively, the fast-fading channels between the BS and the $u$th user (link BS-U), the $u$th IRS and the $u$th user (link IRS-U), and the BS and the $u$th IRS (link BS-IRS), from the polarization $p$ to the polarization $q$, in which $p,q\in \{v,h\}$, with $\chi^{\text{\tiny BS-U}}$ and $\chi^{\text{\tiny BS-IRS}} \in[0,1]$ denoting the inverse of the cross-polar discrimination parameter (iXPD) that measures the power leakage between polarizations in the links BS-U and BS-IRS. Moreover, $\frac{1}{\sqrt{2}}$ is an energy normalization factor, and $\zeta_{kgu}^\text{\tiny BS-U}$, $\zeta_{kgu}^\text{\tiny IRS-U}$, and $\zeta_{kgu}^\text{\tiny BS-IRS}$ represents the large-scale fading coefficients for the links BS-U, IRS-U, and BS-IRS, respectively. Observe that we model depolarization in the links BS-U and BS-IRS, but not in the link IRS-U. This means that only negligible power leaks between polarizations in the channels between the IRSs and users. Furthermore, we assume that $\mathbf{\bar{D}}_{kgu}^{ pq}$, and $\mathbf{\bar{G}}_{kgu}^{ pq}$ are correlated. On the other hand, we model $\mathbf{\bar{S}}_{kgu}^{ pq}$ as a full rank channel matrix. The covariance matrices of the links BS-IRS and BS-U can be calculated, respectively, as $\mathbf{R}^{\text{\tiny BS-IRS}}_{k} = \zeta_{kgu}^\text{\tiny BS-IRS}(\chi^{\text{\tiny BS-IRS}}+1) \mathbf{I}_2 \otimes \mathbf{R}_k$, and $\mathbf{R}^{\text{\tiny BS-U}}_{k} = \zeta_{kgu}^\text{\tiny BS-U}(\chi^{\text{\tiny BS-U}}+1)
\mathbf{I}_2 \otimes \mathbf{R}_k$, where $\mathbf{R}_k$ is the covariance matrix observed in each polarization, with rank denoted by $r_k$. Note that, we have assumed that the links BS-U and BS-IRS share the same covariance matrix $\mathbf{R}_k$. This is valid for the scenario where both IRS and users are located within the same cluster of scatterers. Recalling the Karhunen–Loève representation \cite{ref1}, \eqref{eq:ch0} can be rewritten as
\begin{align}\label{eq:ch1} 
    \mathbf{H}^H_{kgu} &= 
    \text{\footnotesize $ \left(\setlength{\arraycolsep}{2pt} \renewcommand*{\arraystretch}{1.4}
    \begin{bmatrix}
        \mathbf{S}_{kgu}^{ vv} &
        \mathbf{0}_{L,\frac{N}{2}} \\
        \mathbf{0}_{L,\frac{N}{2}} &
        \mathbf{S}_{kgu}^{ hh}
    \end{bmatrix}^H     
    \begin{bmatrix}
        \bm{\Phi}_{kgu}^{vv} & \bm{\Phi}_{kgu}^{hv} \\
        \bm{\Phi}_{kgu}^{vh} & \bm{\Phi}_{kgu}^{hh}
    \end{bmatrix}
    \begin{bmatrix}
        \mathbf{G}_{kgu}^{ v v} &
        \mathbf{G}_{kgu}^{ h v} \\
        \mathbf{G}_{kgu}^{ v h} &
        \mathbf{G}_{kgu}^{ h h}
    \end{bmatrix} \right. $} \nonumber\\[-1mm]
    &+ \text{\scriptsize $\left. 
    \begin{bmatrix}
        \mathbf{D}_{kgu}^{ v v} &
        \mathbf{D}_{kgu}^{ v h} \\
        \mathbf{D}_{kgu}^{ h v} &
        \mathbf{D}_{kgu}^{ h h}
    \end{bmatrix}^H \right) $ \footnotesize$\left(\mathbf{I}_2 \kron \left( \mathbf{\Lambda}_{k}^{\frac{1}{2}} \mathbf{U}^H_{k} \right) \right)$}  \nonumber\\[-1mm]
    & =  \left(\mathbf{S}_{kgu}^H
      \bm{\Theta}_{kgu}
    \mathbf{G}_{kgu} + \mathbf{D}_{kgu}^H \right)  \left(\mathbf{I}_2 \kron \left( \mathbf{\Lambda}_{k}^{\frac{1}{2}} \mathbf{U}^H_{k} \right) \right),
\end{align}
where $\mathbf{\Lambda}_{k} \in \mathbb{R}^{r_k^\star \times r_k^\star}_{>0}$ is a diagonal matrix that collects $r_k^\star$ nonzero eigenvalues of $\mathbf{R}_k$, sorted in descending order, $\mathbf{U}_{k} \in \mathbb{C}^{\frac{M}{2} \times r_k^\star}$ is a unitary matrix containing the first $r_k^\star$ left eigenvectors of $\mathbf{R}_k$, corresponding to the eigenvalues in $\mathbf{\Lambda}_{k}$, $\mathbf{S}_{kgu}^{ pq} \in \mathbb{C}^{L\times\frac{N}{2}}$ is the full rank channel matrix of the link IRS-U, and $\mathbf{D}_{kgu}^{ pq} \in \mathbb{C}^{r_k^\star \times \frac{N}{2}}$ and $\mathbf{G}_{kgu}^{ pq} \in \mathbb{C}^{L\times r_k^\star }$ represent, respectively, the reduced-dimension fast-fading channels of the links BS-U and BS-IRS, from the polarization $p$ to the polarization $q$, with $p,q\in \{v,h\}$, whose entries follow the complex Gaussian distribution with zero mean and unit variance. For notation simplicity, the iXPD, the large scale fading coefficients, and the factor $\frac{1}{\sqrt{2}}$ have been absorbed in the corresponding channel matrices.

The $u$th user in the $g$th group within the $k$th cluster receives the following signal
\begin{align}\label{eq04}
\mathbf{y}_{kgu} &= \left(\mathbf{S}_{kgu}^H
    \bm{\Theta}_{kgu}
    \mathbf{G}_{kgu} + \mathbf{D}_{kgu}^H \right)  \left(\mathbf{I}_2 \kron \left( \mathbf{\Lambda}_{k}^{\frac{1}{2}} \mathbf{U}^H_{k} \right) \right)
 \nonumber\\
 &\times \sum_{m=1}^{K} \mathbf{P}_m\sum_{n=1}^{G} \sum_{i=1}^{U} \mathbf{v}_{mni} \alpha_{mni}{x}_{mni} + \renewcommand*{\arraystretch}{1}\begin{bmatrix} \mathbf{n}^v_{kgu} \\ \mathbf{n}^h_{kgu} \end{bmatrix}, 
\end{align}
where $\mathbf{n}^p_{kgu} \in \mathbb{C}^{\frac{N}{2}\times 1}$ is the noise vector observed at polarization $p\in \{v,h\}$, whose entries follow the complex Gaussian distribution with zero mean and variance $\sigma_n$.

\section{Precoding, IRS Optimization, and Reception}

\subsection{Spatial interference cancellation}\label{ssec:prec}
The precoding matrix $\mathbf{P}_k$ is intended to remove the interference of different spatial clusters. From the signal model in \eqref{eq04}, it is clear that this objective can be accomplished if {\small $\left[\mathbf{I}_2 \otimes \left( \mathbf{\Lambda}_{k}^{\frac{1}{2}} \mathbf{U}^H_{k} \right) \right]  \mathbf{P}_k = \mathbf{0}, \forall k'\neq k$}. Therefore, $\mathbf{P}_k$ can be computed from the null space of the matrix {\small $\bm{\Omega}_k = [\mathbf{U}_{1}, \cdots, \mathbf{U}_{k-1}, \mathbf{U}_{k+1}, \cdots, \mathbf{U}_{K}] \in \mathbb{C}^{\frac{M}{2} \times \sum_{k'\neq k} r^\star_{k'}}$}. To this end, let 
$\mathbf{\Tilde{U}}^{(0)}_k \in \mathbb{C}^{\frac{M}{2}\times \frac{M}{2} - \sum_{k'\neq k} r^\star_{k'}}$ be a unitary matrix composed by the left eigenvectors of $\bm{\Omega}_k$ associated with its last $\frac{M}{2} - \sum_{k'\neq k} r^\star_{k'}$ vanishing eigenvalues. 
Then, to focus the transmissions to the cluster of interest, we match $\mathbf{P}_k$ to the dominant eigenmodes of the matrix {\small $\mathbf{\Pi}_k = \mathbf{I}_2 \otimes  \left[\left( \mathbf{\Tilde{U}}^{(0)}_k \right)^H  \left(\mathbf{U}_{k}  \mathbf{\Lambda}_{k}^{\frac{1}{2}} \right) \right]$}. This can be accomplished by multiplying $\mathbf{\Tilde{U}}^{(0)}_k$ by a unitary matrix constructed from the dominant eigenvectors of the covariance matrix {\small $\mathbf{\Pi}_k (\mathbf{\Pi}_k)^H  = \mathbf{I}_2 \otimes \left[\left( \mathbf{\Tilde{U}}^{(0)}_k \right)^H \mathbf{R}_k \mathbf{\Tilde{U}}^{(0)}_k \right] = \mathbf{I}_2 \otimes \mathbf{\tilde{\Xi}}_{k}$}. To be more specific, by representing the left eigenvectors of $\mathbf{\tilde{\Xi}}_{k}$ by $\mathbf{\bar{U}}_k = \begin{bmatrix} \mathbf{\bar{U}}^{(1)}_k & \mathbf{\bar{U}}^{(0)}_k \end{bmatrix}$, with $\mathbf{\bar{U}}^{(1)}_k \in \mathbb{C}^{\left( \frac{M}{2} - \sum_{k'\neq k} r^\star_{k'} \right)\times \frac{\bar{M}}{2} } $ collecting the first $\frac{\bar{M}}{2}$ columns of $\mathbf{\bar{U}}_k$, the desired precoding matrix can be computed by $\mathbf{P}_k = \mathbf{I}_2 \otimes \left( \mathbf{\Tilde{U}}^{(0)}_k \mathbf{\bar{U}}^{(1)}_k \right) = \mathbf{I}_2 \otimes \mathbf{\Tilde{P}}_k \in \mathbb{C}^{ M \times \bar{M}}$, in which, due to the dimensions of $\mathbf{\tilde{U}}_k$ and {$\mathbf{\bar{U}}_k$}, the constraints {\small $K \leq \bar{M} \leq \left( M - 2\sum_{k'\neq k} r^\star_{k'} \right)$} and {\small $\bar{M} \leq 2r^\star_{k}$} must be satisfied. \vspace{-3mm}

\begin{figure*}[!t]
	\normalsize
	\setcounter{MYtempeqncnt}{\value{equation}}
	\setcounter{equation}{9}
	\begin{align}\label{sigtop}
    \mathbf{y}_{gu} &= \text{\footnotesize $\renewcommand*{\arraystretch}{1.6} \left(\begin{bmatrix}
            \left[ (\mathbf{S}_{gu}^{ vv})^H
        \bm{\Phi}_{gu}^{vv}
        \mathbf{G}_{gu}^{ vv}
        +
        (\mathbf{S}_{gu}^{ vv})^H
        \bm{\Phi}_{gu}^{hv}
        \mathbf{G}_{gu}^{ vh} \right] & \left[ (\mathbf{S}_{gu}^{ vv})^H
        \bm{\Phi}_{gu}^{vv}
        \mathbf{G}_{gu}^{ hv}
        +
        (\mathbf{S}_{gu}^{ vv})^H
        \bm{\Phi}_{gu}^{hv}
        \mathbf{G}_{gu}^{ hh} \right] \\
         \left[ (\mathbf{S}_{gu}^{ hh})^H
        \bm{\Phi}_{gu}^{vh}
        \mathbf{G}_{gu}^{ vv}
        +
        (\mathbf{S}_{gu}^{ hh})^H
        \bm{\Phi}_{gu}^{hh}
        \mathbf{G}_{gu}^{ vh} \right] & \left[ (\mathbf{S}_{gu}^{ hh})^H
        \bm{\Phi}_{gu}^{vh}
        \mathbf{G}_{gu}^{ hv}
        +
        (\mathbf{S}_{gu}^{ hh})^H
        \bm{\Phi}_{gu}^{hh}
        \mathbf{G}_{gu}^{ hh} \right]
    \end{bmatrix} + 
    \renewcommand*{\arraystretch}{1.6}
\begin{bmatrix}
        (\mathbf{D}_{gu}^{ v v})^H &
        (\mathbf{D}_{gu}^{ h v})^H \\
        (\mathbf{D}_{gu}^{ v h})^H &
        (\mathbf{D}_{gu}^{ h h})^H
    \end{bmatrix}\right)$} \nonumber\\
    & \text{\footnotesize $\times \renewcommand*{\arraystretch}{1.4}
    \begin{bmatrix}
        \mathbf{\Lambda}^{\frac{1}{2}} \mathbf{U}^H \mathbf{\Tilde{P}} & \mathbf{0}_{\frac{M}{2}, \frac{\bar{M}}{2}} \\
        \mathbf{0}_{\frac{M}{2}, \frac{\bar{M}}{2}} &
        \mathbf{\Lambda}^{\frac{1}{2}} \mathbf{U}^H \mathbf{\Tilde{P}}
    \end{bmatrix}
  \sum_{n=1}^{G} \sum_{i=1}^{U} 
  \begin{bmatrix}
    \mathbf{v}^v_{ni} \\
    \mathbf{v}^h_{ni}
    \end{bmatrix}
  \alpha_{ni}{x}_{ni} + \begin{bmatrix} \mathbf{n}^v_{gu} \\ \mathbf{n}^h_{gu} \end{bmatrix}.$}
  \end{align}
	\setcounter{equation}{\value{MYtempeqncnt}+1}
	\vspace*{-3pt}
	\hrulefill
	\vspace*{-14pt}
\end{figure*}

\subsection{Polarization assignment and formation of subsets} \label{passi}
First, the BS sorts the users within each group in ascending order based on their large-scale fading coefficients observed in the link BS-U, such that $\zeta^{\text{\tiny BS-U}}_{kg1} < \zeta^{\text{\tiny BS-U}}_{kg2} < \cdots < \zeta^{\text{\tiny BS-U}}_{kgU}$. Then, by assuming that $U$ is an even number, users associated with odd indexes are assigned to the vertical polarization, and users associated with even indexes to the horizontal polarization, resulting in two disjoint subsets, the vertical subset $\mathcal{U}^{v} = \{1, 3,\cdots, U - 1\}$, containing $U^v = U/2$ users, and the horizontal subset $\mathcal{U}^{h} = \{2, 4, \cdots, U\}$, containing $U^h = U - U^v = U/2$ users. In order to implement this strategy, for $1 \leq g \leq G$ and $1 \leq u \leq U$, the BS employs the following precoding vector
\begin{align}\label{pvec}
    \mathbf{v}_{kgu} = \renewcommand*{\arraystretch}{1}\begin{bmatrix}
    \mathbf{v}^v_{kgu} \\
    \mathbf{v}^h_{kgu}
    \end{bmatrix} =
    \renewcommand*{\arraystretch}{1.2}
    \begin{bmatrix} 
    \left[\mathbf{0}_{1,g-1}, \mathbf{1}_{\mathcal{U}^{v}}(u), \mathbf{0}_{1,\frac{\bar{M}}{2}-g}\right]^T \\
    \left[\mathbf{0}_{1,g-1}, \mathbf{1}_{\mathcal{U}^{h}}(u), \mathbf{0}_{1,\frac{\bar{M}}{2}-g}\right]^T
    \end{bmatrix},
\end{align}
where $\mathbf{1}_{\mathcal{A}}(i)$ is the indicator function of a subset $\mathcal{A}$, which results $1$ if $i \in \mathcal{A}$, and $0$ if $i \notin \mathcal{A}$.

\subsection{IRS optimization}
Since the inter-cluster interference has been addressed, we can drop the cluster subscript and simplify the signal in \eqref{eq04} as in \eqref{sigtop}, on the top of the next page.

As can be observed in \eqref{sigtop}, the symbols intended to the subsets assigned to the vertical polarization propagate through the channels modeled by the left blocks of the channel matrices, while the symbols for subsets assigned to the horizontal polarization propagate through the right blocks. Therefore, the IRSs of users assigned to the vertical polarization should be optimized to null out the right channel blocks, and the IRSs for users assigned to the horizontal polarization should null out the left channel blocks. More specifically, we aim to achieve in subsets assigned to the vertical polarization:
\begin{align}\label{objvert}
    &\text{\footnotesize $\renewcommand*{\arraystretch}{1}\begin{bmatrix}
       (\mathbf{S}_{gu}^{ vv})^H
        \bm{\Phi}_{gu}^{vv}
        \mathbf{G}_{gu}^{ hv}
        \hspace{-.5mm}+\hspace{-.5mm}
        (\mathbf{S}_{gu}^{ vv})^H
        \bm{\Phi}_{gu}^{hv}
        \mathbf{G}_{gu}^{ hh} \\
       (\mathbf{S}_{gu}^{ hh})^H
        \bm{\Phi}_{gu}^{vh}
        \mathbf{G}_{gu}^{ hv}
        \hspace{-.5mm}+\hspace{-.5mm}
        (\mathbf{S}_{gu}^{ hh})^H
        \bm{\Phi}_{gu}^{hh}
        \mathbf{G}_{gu}^{ hh}
    \end{bmatrix} \hspace{-.8mm} + \hspace{-.8mm}
    \begin{bmatrix}
        (\mathbf{D}_{gu}^{ h v})^H \\
        (\mathbf{D}_{gu}^{ h h})^H
    \end{bmatrix}
    \hspace{-1mm} \approx \hspace{-1mm} \begin{bmatrix}
        \mathbf{0}_{\frac{N}{2},r^\star_k} \\
        \mathbf{0}_{\frac{N}{2},r^\star_k}
    \end{bmatrix}$}\hspace{-.5mm},
\end{align}
and in subsets assigned to the horizontal polarization:
\begin{align}
    &\text{\footnotesize $\renewcommand*{\arraystretch}{1}\begin{bmatrix}
        (\mathbf{S}_{gu}^{ vv})^H
        \bm{\Phi}_{gu}^{vv}
        \mathbf{G}_{gu}^{ vv}
        \hspace{-.5mm}+\hspace{-.5mm}
        (\mathbf{S}_{gu}^{ vv})^H
        \bm{\Phi}_{gu}^{hv}
        \mathbf{G}_{gu}^{ vh}
      \\
        (\mathbf{S}_{gu}^{ hh})^H
        \bm{\Phi}_{gu}^{vh}
        \mathbf{G}_{gu}^{ vv}
        \hspace{-.5mm}+\hspace{-.5mm}
        (\mathbf{S}_{gu}^{ hh})^H
        \bm{\Phi}_{gu}^{hh}
        \mathbf{G}_{gu}^{ vh}
    \end{bmatrix} \hspace{-.5mm}+\hspace{-.5mm}
    \begin{bmatrix}
        (\mathbf{D}_{gu}^{ v v})^H \\
        (\mathbf{D}_{gu}^{ v h})^H
    \end{bmatrix}
    \hspace{-1mm}\approx\hspace{-1mm} \begin{bmatrix}
        \mathbf{0}_{\frac{N}{2},r^\star_k} \\
        \mathbf{0}_{\frac{N}{2},r^\star_k}
    \end{bmatrix}$}\hspace{-.5mm}.
\end{align}

Note that, by mitigating the transmissions originated from the interfering polarization, we can transform depolarization phenomena into an advantage. More specifically, this strategy should enable users to receive their intended messages, transmitted from a single polarization (or vertical, or horizontal), in both receive polarizations. In other words, the proposed scheme enables polarization diversity. 

Due to space constraints, we focus on the optimization for IRSs of vertical subsets. Based on \eqref{objvert}, the reflecting coefficients for users assigned to the vertical polarization can be optimized by solving the following problem
\begin{subequations}\label{p1}
\begin{align}
        & \text{\small $\underset{\bm{\Phi}_{gu}^{hv}, \bm{\Phi}_{gu}^{hh} }{\underset{\bm{\Phi}_{gu}^{vv}, \bm{\Phi}_{gu}^{vh}}{\min}}
        \hspace{-1.5mm}$}\hspace{.8mm}\text{\scriptsize $\left\| \renewcommand*{\arraystretch}{1}
        \begin{bmatrix}
       (\mathbf{S}_{gu}^{ vv})^H
        \bm{\Phi}_{gu}^{vv}
        \mathbf{G}_{gu}^{ hv} \\
       (\mathbf{S}_{gu}^{ hh})^H
        \bm{\Phi}_{gu}^{vh}
        \mathbf{G}_{gu}^{ hv}
    \end{bmatrix} \hspace{-1.5mm} + \hspace{-1.5mm}\begin{bmatrix}
        (\mathbf{S}_{gu}^{ vv})^H
        \bm{\Phi}_{gu}^{hv}
        \mathbf{G}_{gu}^{ hh} \\
        (\mathbf{S}_{gu}^{ hh})^H
        \bm{\Phi}_{gu}^{hh}
        \mathbf{G}_{gu}^{ hh}
    \end{bmatrix} \hspace{-1.5mm}+\hspace{-1.5mm}
    \begin{bmatrix}
        (\mathbf{D}_{gu}^{ h v})^H \\
        (\mathbf{D}_{gu}^{ h h})^H
    \end{bmatrix}
          \right\|^2$} \\[-1mm]
    &\quad\text{ s.t.} \hspace{2mm} \text{ $|\omega^{pq}_{gu,l}|^2 \leq 1, \hspace{2mm}  \forall l \in [1, L], \forall p,q \in \{v,h\},$} \label{p12b} \\[-1mm]
    &\quad \quad \text{ $\bm{\Phi}_{gu}^{vv}, \bm{\Phi}_{gu}^{vh}, \bm{\Phi}_{gu}^{hv}, \bm{\Phi}_{gu}^{hh}$} \text{ diagonal.}
\end{align}
\end{subequations}
As one can notice, due to the element-wise quadratic constraint and the diagonal matrices constraint, it becomes difficult to solve \eqref{p1} in its current form. To overcome this challenge, we transform \eqref{p1} in an equivalent tractable problem. Using the identity {\small $(\mathbf{C}^T\hspace{-1mm} \odot\hspace{-.2mm} \mathbf{A}) \textit{vecd}\{\mathbf{B}\}\hspace{-1mm} = \hspace{-1mm}\textit{vec}\{ \mathbf{A}\mathbf{B}\mathbf{C}\} $} \cite{Brewer78}, we define:
\begin{align*}
  &\text{\footnotesize $\bm{\theta}^{pq}_{gu} = \textit{vecd}\{\bm{\Phi}^{pq}_{gu}\},
  \hspace{3mm}
  \mathbf{d}^{hv}_{gu} = \textit{vec}\left\{(\mathbf{D}_{gu}^{ h v})^H \right\}, \hspace{3mm} \mathbf{d}^{hh}_{gu} = \textit{vec}\left\{(\mathbf{D}_{gu}^{ h h})^H \right\}$}
  \\[-.5mm]
  &\text{\footnotesize $\mathbf{K}^{hv,vv}_{gu} = [(\mathbf{G}_{gu}^{ hv})^T \kr (\mathbf{S}_{gu}^{ vv})^H], \hspace{3mm} \mathbf{K}^{hh,vv}_{gu} = [(\mathbf{G}_{gu}^{ hh})^T \kr (\mathbf{S}_{gu}^{ vv})^H]  $}\\[-.5mm]
   &  \text{\footnotesize $\mathbf{K}^{hv,hh}_{gu} = [(\mathbf{G}_{gu}^{ hv})^T \kr (\mathbf{S}_{gu}^{ hh})^H], \hspace{3mm} \mathbf{K}^{hh,hh}_{gu} = [(\mathbf{G}_{gu}^{ hh})^T \kr (\mathbf{S}_{gu}^{ hh})^H]$}.
\end{align*}
Then, \eqref{p1} is transformed into the following two sub-problems\vspace{-3mm}
\begin{subequations}\label{p2}
\begin{align}
        &\text{\small $\underset{\bm{\theta}^{vv}_{gu}, \bm{\theta}^{hv}_{gu}}{\min}
         \left\|\setlength{\arraycolsep}{1pt}
        \begin{bmatrix}
       \mathbf{K}^{hv,vv}_{gu} &
        \mathbf{K}^{hh,vv}_{gu} \end{bmatrix} \left[(\bm{\theta}^{vv}_{gu})^T, (\bm{\theta}^{hv}_{gu})^T\right]^T \hspace{-2mm}  +
    \mathbf{d}^{hv}_{gu}
        \right\|^2$} \label{p2a}\\[-2mm]
    &\quad\text{s.t.} \hspace{2mm} \text{\small $\left\| \left[(\bm{\theta}^{vv}_{gu})^T, (\bm{\theta}^{hv}_{gu})^T \right]^T \right\|^2_{\infty} \leq 1,$}\label{p2b}
\end{align}
\end{subequations}\vspace{-3mm}
\begin{subequations}\label{p3}
\begin{align}
        &\text{\small $\underset{\bm{\theta}^{vh}_{gu}, \bm{\theta}^{hh}_{gu} }{\min}\hspace{-1mm}
         \left\| \setlength{\arraycolsep}{1pt}
        \begin{bmatrix}
       \mathbf{K}^{hv,hh}_{gu} &
       \mathbf{K}^{hh,hh}_{gu} \end{bmatrix} \left[(\bm{\theta}^{vh}_{gu})^T, (\bm{\theta}^{hh}_{gu})^T \right]^T  \hspace{-2mm}  +
    \mathbf{d}^{hh}_{gu} 
        \right\|^2$} \label{p3a}\\[-2mm]
    &\quad\text{s.t.} \hspace{2mm} \text{\small $\left\| \left[(\bm{\theta}^{vh}_{gu})^T, (\bm{\theta}^{hh}_{gu})^T \right]^T \right\|^2_{\infty} \leq 1.$}\label{p3b}
\end{align}
\end{subequations}
Before we can solve the problems above, let us denote
{\footnotesize $\mathbf{\bar{K}}_{gu} \hspace{-1mm}=\hspace{-1mm} \setlength{\arraycolsep}{1pt}\begin{bmatrix} \mathbf{K}^{hv,vv}_{gu} & \mathbf{K}^{hh,vv}_{gu} \end{bmatrix}$}, {\footnotesize $\mathbf{\bar{C}}_{gu} \hspace{-1mm}=\hspace{-1mm} \mathbf{\bar{K}}_{gu}^H\mathbf{\bar{K}}_{gu}$}, and
{\footnotesize$\mathbf{\Tilde{K}}_{gu} \hspace{-1mm}=\hspace{-1mm} \setlength{\arraycolsep}{1pt} \begin{bmatrix} \mathbf{K}^{hv,hh}_{gu} & \mathbf{K}^{hh,hh}_{gu} \end{bmatrix}$}, {\footnotesize$\mathbf{\Tilde{C}}_{gu} = \mathbf{\Tilde{K}}_{gu}^H\mathbf{\Tilde{K}}_{gu}$}, and rewrite the left-hand side of the constraints in \eqref{p2b} and \eqref{p3b}, respectively, as 
{\footnotesize $\left\| \left[(\bm{\theta}^{vv}_{gu})^T \hspace{0mm} , (\bm{\theta}^{hv}_{gu})^T \right]^T \right\|^2_{\infty} \hspace{-1mm}=$ $\left[(\bm{\theta}^{vv}_{gu})^H \hspace{-1mm}, (\bm{\theta}^{hv}_{gu})^H \right] \hspace{0mm} \mathbf{B}_l \hspace{0mm} \left[(\bm{\theta}^{vv}_{gu})^T \hspace{-1mm}, (\bm{\theta}^{hv}_{gu})^T \right]^T$ }, and
{\footnotesize $\left\| \left[(\bm{\theta}^{vh}_{gu})^T \hspace{-2mm} , (\bm{\theta}^{hh}_{gu})^T \right]^T \right\|^2_{\infty} \hspace{-3mm}=$ $\left[(\bm{\theta}^{vh}_{gu})^H \hspace{-1mm}, (\bm{\theta}^{hh}_{gu})^H \right] \hspace{-1mm}\mathbf{B}_l \hspace{-1mm}\left[(\bm{\theta}^{vh}_{gu})^T\hspace{-1mm}, (\bm{\theta}^{hh}_{gu})^T \right]^T \hspace{-1mm}$}, where {\footnotesize$\mathbf{B}_l = \textit{diag}\{\mathbf{e}_l\}, l=1,\cdots, L$}, with $\mathbf{e}_l$ representing the standard basis vector that contains 1 in the $l$th position and zeros elsewhere.
Then, by expanding the objective functions in \eqref{p2a} and \eqref{p3a}, we obtain\vspace{-3mm}
\begin{subequations}\label{p4}
\begin{align}
        \text{\footnotesize$\underset{\bm{\theta}^{vv}_{gu}, \bm{\theta}^{hv}_{gu}}{\min}$} \hspace{-1mm}
         & \text{\footnotesize $\left\{ \renewcommand{\arraystretch}{0.9}
         \begin{bmatrix}
                \bm{\theta}^{vv}_{gu} \\
                \bm{\theta}^{hv}_{gu}
         \end{bmatrix}^H \hspace{-2mm}
         \mathbf{\bar{C}}_{gu} \begin{bmatrix}
                \bm{\theta}^{vv}_{gu} \\
                \bm{\theta}^{hv}_{gu}
         \end{bmatrix} + 2\Re\hspace{-.6mm}\left\{(\mathbf{d}^{hv}_{gu})^H\mathbf{\bar{K}}_{gu}\hspace{-1mm} \begin{bmatrix}
                \bm{\theta}^{vv}_{gu} \\
                \bm{\theta}^{hv}_{gu}
         \end{bmatrix} \right\} \hspace{-.6mm} + \hspace{-.6mm} (\mathbf{d}^{hv}_{gu})^H\mathbf{d}^{hv}_{gu}\right\}$}
         \\[-4mm]
    \quad\text{\footnotesize s.t.} &\hspace{1mm} \text{\footnotesize $\renewcommand{\arraystretch}{0.9} \begin{bmatrix}
                \bm{\theta}^{vv}_{gu} \\
                \bm{\theta}^{hv}_{gu}
         \end{bmatrix}^H\hspace{-2mm} \mathbf{B}_l \begin{bmatrix}
                \bm{\theta}^{vv}_{gu} \\
                \bm{\theta}^{hv}_{gu}
         \end{bmatrix} \leq 1,$}\label{p4b}
\end{align}
\end{subequations}\vspace{-1mm}
\begin{subequations}\label{p5}
\begin{align}
        \text{\footnotesize $\underset{\bm{\theta}^{vh}_{gu}, \bm{\theta}^{hh}_{gu}}{\min}$}&\hspace{-1mm}
         \text{\footnotesize $\left\{
         \renewcommand{\arraystretch}{0.9}
         \begin{bmatrix}
                \bm{\theta}^{vh}_{gu} \\
                \bm{\theta}^{hh}_{gu}
         \end{bmatrix}^H \hspace{-2mm}
         \mathbf{\Tilde{C}}_{gu} \begin{bmatrix}
                \bm{\theta}^{vh}_{gu} \\
                \bm{\theta}^{hh}_{gu}
         \end{bmatrix} + 2\Re\hspace{-.6mm}\left\{\hspace{-.6mm}(\mathbf{d}^{hh}_{gu})^H\mathbf{\Tilde{K}}_{gu}\hspace{-1mm} \begin{bmatrix}
                \bm{\theta}^{vh}_{gu} \\
                \bm{\theta}^{hh}_{gu}
         \end{bmatrix} \right\} \hspace{-.6mm} + \hspace{-.6mm} (\mathbf{d}^{hh}_{gu})^H\mathbf{d}^{hh}_{gu}\right\}$}
         \\[-4mm]
    \quad\text{\footnotesize s.t.}& \hspace{1mm} \text{\footnotesize $\renewcommand{\arraystretch}{0.9}
         \begin{bmatrix}
                \bm{\theta}^{vh}_{gu} \\
                \bm{\theta}^{hh}_{gu}
         \end{bmatrix}^H \hspace{-2mm} \mathbf{B}_l \begin{bmatrix}
                \bm{\theta}^{vh}_{gu} \\
                \bm{\theta}^{hh}_{gu}
         \end{bmatrix} \leq 1$}.\label{p5b}
\end{align}
\end{subequations}

Problems \eqref{p4} and \eqref{p5} are quadractically constrained quadratic problems. Since the entries of $\mathbf{\bar{K}}_{gu}$ and $\mathbf{\Tilde{K}}_{gu}$ are independent complex Gaussian random variables, $\mathbf{\bar{C}}_{gu}$ and $\mathbf{\Tilde{C}}_{gu}$ will be positive semidefinite matrices. Furthermore, since $\mathbf{z}^H\mathbf{B}_l \mathbf{z} = |[\mathbf{z}]_{l}|^2 \geq 0, \hspace{0mm} \forall \mathbf{z} \in \mathbb{C}^{L\times 1}$, $\mathbf{B}_l$ is also positive semidefinite. As a result, \eqref{p4} and \eqref{p5} are convex and, consequently, have global optimal solutions that can be efficiently computed via interior-points methods in polynomial time \cite{Luo2010}. \vspace{-3mm}

\subsection{Signal reception}
For the sake of simplicity, hereinafter the links BS-IRS-U and BS-U are absorbed into a single channel matrix, and \eqref{eq:ch1} is rewritten in a more compact structure, as follows
\begin{align}\label{eq:ch24}
    \mathbf{H}^H_{gu} = \renewcommand*{\arraystretch}{1} \begin{bmatrix}
        \mathbf{\Tilde{H}}_{gu}^{v v} &
        \mathbf{\Tilde{H}}_{gu}^{v h} \\
        \mathbf{\Tilde{H}}_{gu}^{h v} &
        \mathbf{\Tilde{H}}_{gu}^{h h}
    \end{bmatrix}^H,
\end{align}
where $\mathbf{\Tilde{H}}_{gu}^{pq}$ accounts for both direct and reflected transmissions that depart the BS from polarization $p$ and arrive at the user's devices on polarization $q$, with $p,q \in \{v,h\}$, e.g., the effective vertical-to-vertical channel matrix is defined by {\footnotesize $\mathbf{\Tilde{H}}_{gu}^{vv} =  \mathbf{U}_{k}  \mathbf{\Lambda}_{k}^{\frac{1}{2}} \left[ (\mathbf{S}_{gu}^{ vv})^H \bm{\Phi}_{gu}^{vv} \mathbf{G}_{gu}^{ vv}\right.$ + $\left.(\mathbf{S}_{gu}^{ vv})^H \bm{\Phi}_{gu}^{hv} \mathbf{G}_{gu}^{ vh} \right]^H$ $ + \hspace{1mm} \mathbf{U}_{k}  \mathbf{\Lambda}_{k}^{\frac{1}{2}} \mathbf{D}_{gu}^{ vv} $}. With this notation, the signal in \eqref{sigtop} can be simplified to 
\begin{align}\label{sigrec1}
\mathbf{y}_{gu} \hspace{-1mm}=\hspace{-1mm} \text{\footnotesize $ \renewcommand*{\arraystretch}{1}\setlength{\arraycolsep}{2pt}
  \begin{bmatrix}
        (\mathbf{\Tilde{H}}_{gu}^{v v})^H\mathbf{\Tilde{P}} &
        (\mathbf{\Tilde{H}}_{gu}^{h v})^H\mathbf{\Tilde{P}} \\
        (\mathbf{\Tilde{H}}_{gu}^{v h})^H \mathbf{\Tilde{P}} &
        (\mathbf{\Tilde{H}}_{gu}^{h h})^H \mathbf{\Tilde{P}}
    \end{bmatrix} \hspace{-1mm} \sum_{n=1}^{G}  \sum_{i=1}^{U} 
  \begin{bmatrix}
    \mathbf{v}^v_{ni} \\
    \mathbf{v}^h_{ni}
    \end{bmatrix}
  \alpha_{ni}{x}_{ni}
    + \begin{bmatrix} \mathbf{n}^v_{gu} \\ \mathbf{n}^h_{gu} \end{bmatrix} $}.
\end{align}

Then, in order to explain our detection strategy, we focus on subsets assigned to the vertical polarization. Remember that the IRSs of users assigned to the vertical polarization are optimized to mitigate all transmissions originated at the BS from the horizontal polarization. Therefore, by relying on the effectiveness of the IRS, we exploit the left blocks of the channel matrix in \eqref{sigrec1} to construct our detection matrix. More specifically, in order to remove the remaining interference from other subsets also assigned to the vertical polarization, the $u$th user exploits the virtual channels $\mathbf{\underline{H}}_{gu}^{v v} = (\mathbf{\Tilde{H}}_{gu}^{v v})^H\mathbf{\Tilde{P}}$ and $\mathbf{\underline{H}}_{gu}^{v h} = (\mathbf{\Tilde{H}}_{gu}^{v h})^H\mathbf{\Tilde{P}}$ to construct the detection matrices
$\mathbf{H}^{\dagger v}_{gu} = [(\mathbf{\underline{H}}_{gu}^{v v})^H \mathbf{\underline{H}}_{gu}^{v v}]^{-1}(\mathbf{\underline{H}}_{gu}^{v v})^H$ and $\mathbf{H}^{\dagger h}_{gu} = [(\mathbf{\underline{H}}_{gu}^{v h})^H \mathbf{\underline{H}}_{gu}^{v h}]^{-1}(\mathbf{\underline{H}}_{gu}^{v h})^H$, in which it is assumed that $N\geq \bar{M}$. Then, after multipliying  \eqref{sigrec1} by these matrices, the $u$th user obtains the following data vector
\begin{align}\label{sigrec2}
\mathbf{\hat{x}}_{gu} &=\renewcommand*{\arraystretch}{1} \begin{bmatrix} \mathbf{\hat{x}}^{v}_{gu} \\ \mathbf{\hat{x}}^{h}_{gu} \end{bmatrix} =  \setlength{\arraycolsep}{3pt}
  \begin{bmatrix}
        \mathbf{x}^v +
        \mathbf{H}^{\dagger v}_{gu}\mathbf{\underline{H}}_{gu}^{h v} \mathbf{x}^h \\
        \mathbf{x}^v +
        \mathbf{H}^{\dagger h}_{gu}\mathbf{\underline{H}}_{gu}^{h h} \mathbf{x}^h
    \end{bmatrix} + \begin{bmatrix} \mathbf{H}^{\dagger v}_{gu}\mathbf{n}^v_{gu} \\ \mathbf{H}^{\dagger h}_{gu} \mathbf{n}^h_{gu} \end{bmatrix},
\end{align}
where, due to the precoding vector in \eqref{pvec}, $\mathbf{x}^v$ is given by
\begin{align}\label{vecv}
\mathbf{x}^v &= \renewcommand*{\arraystretch}{.5} \begin{bmatrix} 
            \sum_{i\in \mathcal{U}^{v}} \alpha_{1i}x_{1i} \vspace{-1mm} \\
            \vdots \vspace{-1mm} \\
            \sum_{i\in \mathcal{U}^{v}} \alpha_{Gi}x_{Gi}
         \end{bmatrix}.
\end{align}

Note in \eqref{sigrec2} that users will obtain in both receive polarizations corrupted replicas of the vector of superimposed symbols that was transmitted by the BS from the vertical polarization.
Therefore, a user within the $g$th vertical subset is able to decode its symbol from the $g$th element of both $\mathbf{\hat{x}}^{v}_{gu}$ and $\mathbf{\hat{x}}^{h}_{gu}$. Inspired by the strategy proposed in \cite{ni3}, the symbols will be decoded from the polarization that renders the highest effective channel gain, denoted in this work as the polarization $\ddot{p}$. As a result, the superimposed symbol recovered by the $u$th user in the $g$th vertical subset before carrying out SIC is given by
\begin{align}\label{datavec1}
    [\mathbf{\hat{x}}^{\ddot{p}}_{gu}]_{g} = \sum_{i\in \mathcal{U}^{v}} \alpha_{gi}x_{gi} +
        [\mathbf{H}^{\dagger \ddot{p}}_{gu}\mathbf{\underline{H}}_{gu}^{h \ddot{p}}\mathbf{x}^h]_{g} + [\mathbf{H}^{\dagger \ddot{p}}_{gu}\mathbf{n}^{\ddot{p}}_{gu}]_{g}.
\end{align}

Users within horizontal subsets employ the same strategy. \vspace{-3mm}

\subsection{SINR analysis}
Recall that users within each subset are sorted in ascending order based on their large scale coefficients. Thus, before the $u$th user in the polarization subset $\mathcal{U}^{p}$, $p\in \{v,h\}$, can retrieve its own message, it carries out SIC to decode the symbol intended for the $m$th weaker user,  $\forall m < u,$ $m\in \mathcal{U}^{p}$, and treats the message to the $n$th stronger user as interference, $\forall n > u, n \in \mathcal{U}^{p}$. Ideally, the symbols intended for weaker users can be perfectly removed by SIC. However, SIC errors are inevitable in practice. Therefore, users suffer from SIC error propagation in our system, and this is modeled as a linear function of the power of decoded symbols, as in \cite{SenaISIC2020}. Then, after all SIC decodings, the $u$th user within the subset $\mathcal{U}^{p}$ in the $g$th group observes the following symbol
\begin{align}\label{sigbsic4}
    {\hat{x}}_{gu} &=\hspace{-1mm} \underset{\text{Desired symbol}}{\underbrace{ 
    \alpha_{gu}x_{gu}
    }} 
    + \hspace{-2mm}\underset{\text{Interference of stronger users}}{\underbrace{
    \sum_{m \in \{a | \hspace{.5mm} a > u, \hspace{.5mm} a\in \mathcal{U}^{p} \}} \hspace{-8mm} \alpha_{gm}x_{gm}
    }}
    + \underset{\text{Residual SIC interference}}{\underbrace{
    \sqrt{\xi} \hspace{-8mm} \sum_{n \in \{b | \hspace{.5mm} b < u, \hspace{.5mm} b\in \mathcal{U}^{p} \}} \hspace{-8mm} \alpha_{gn}x_{gn}
    }} 
   \nonumber\\
   &\hspace{-1mm} + \underset{\text{Polarization interference}}{\underbrace{
    [\mathbf{H}^{\dagger \ddot{p}}_{gu}\mathbf{\underline{H}}_{gu}^{t \ddot{p}} \mathbf{x}^t]_{g}
    }}
    \hspace{-2mm} + \underset{\text{Noise}}{\underbrace{
    [\mathbf{H}^{\dagger \ddot{p}}_{gu}\mathbf{n}^{\ddot{p}}_{gu}]_{g}
    }},
\end{align}
where the superscript $t$ represents the interfering polarization that is defined by $t= h$, if $u\in \mathcal{U}^{v}$, or $t = v$, if $u\in \mathcal{U}^{h}$, and $\xi \in [0,1]$ is the SIC error factor. The signal-to-interference-plus-noise ratio (SINR) observed during each SIC decoding is defined in the following lemma.

\paragraph*{Lemma I} The $u$th user, when decoding the symbol to the $i$th user, $\forall i \leq u$, $i\in \mathcal{U}^{p}$, observes the following SINR
\begin{align}\label{sinreq}
\gamma_{gu}^{i} &= \frac{\rho\mathcal{\ddot{h}}_{gu} \alpha_{gi}^2}{\rho\mathcal{\ddot{h}}_{gu}\mathfrak{I}_{gi} + \rho\mathcal{\ddot{h}}_{gu}\mathfrak{X}_{gu} + 
1},
\end{align}
where $\mathcal{\ddot{h}}_{gu} = \max \{\mathcal{h}^{v}_{gu}, \mathcal{h}^{h}_{gu}\}$, with $\mathcal{h}^{p}_{gu} = [1/\mathbf{H}^{\dagger p}_{gu}(\mathbf{H}^{\dagger p}_{gu})^H]_{gg}$ being the effective channel in polarization $p$, $\mathfrak{X}_{gu} = \left|[\mathbf{H}^{\dagger \ddot{p}}_{gu}\mathbf{\underline{H}}_{gu}^{t \ddot{p}}\mathbf{x}^t]_{g} \right|^2$ is the polarization interference, in which, if $u\in \mathcal{U}^{v}$, $t = h$, and if $u\in \mathcal{U}^{h}$, $t=v$. The symbol $\rho = 1/\sigma_n^{2}$ represents the signal-to-noise ratio (SNR), and $\mathfrak{I}_{gi}$ is the total SIC interference given by
\begin{align}
    \text{\footnotesize $\mathfrak{I}_{gi} \hspace{-1mm}=$}\hspace{-1mm}
    \begin{cases}
        \sum_{m=i+1}^{\max\{\mathcal{U}^{p}\}} \alpha_{gm}^2, \hspace{-2mm} \quad \text{ if } i = \min\{\mathcal{U}^{p}\},\\
        \xi \sum_{n=1}^{i-1} \alpha_{gn}^2, \hspace{-2mm} \quad \text{ if } i = u = \max\{\mathcal{U}^{p}\},\\
        \hspace{-1mm} \sum_{m=i+1}^{\max\{\mathcal{U}^{p}\}} \hspace{-1mm} \alpha_{gm}^2 \hspace{-1mm} + \hspace{-.8mm}\xi \sum_{n=\min\{\mathcal{U}^{p}\}}^{i-1} \hspace{-1mm} \alpha_{gn}^2, \hspace{-2mm} \hspace{1mm} \text{ otherwise.}
    \end{cases}
\end{align}

\textit{Proof:} Please, see Appendix \ref{ap1}.  \hfill\qedsymbol \vspace{-2mm}

\section{Simulation Results and Discussions}
In this section, we use as baseline schemes the classical MIMO-OMA system, where users are served via time division multiple access, and the conventional single-polarized and dual-polarized MIMO-NOMA systems, whose implementation details can be found in \cite{ni3}. For a fair performance comparison, in both single and dual-polarized schemes, we employ at the BS a linear array with $M=90$ transmit antennas. We generate the covariance matrices of the links BS-IRS and BS-U through the one-ring geometrical model \cite{ni3,ref1}, where we consider $K = 4$ spatial clusters, each with $30$~m of radius and located at $120$~m from the BS. In addition, the cluster from which the simulation results are generated is positioned at the azimuth angle of $30^\circ$, and it comprises $G=\bar{M}=4$ groups, each one containing $U=4$ users. In particular, we focus on the first group, where the users $1$, $2$, $3$ and $4$ are located, respectively, at $135$~m, $125$~m, $115$~m, and $105$~m from the BS. A fixed power allocation is adopted, in which we set $\alpha_{1}^2 = 0.4, \alpha_{2}^2 =0.35, \alpha_{3}^2 = 0.2, \alpha_{4}^2 = 0.05$. Moreover, we assume that the distances from the BS to each IRS are the same as that from the BS to its connected user. Under these assumptions, the fading coefficients for the links BS-U and BS-IRS are configured as $\zeta_{u}^\text{\tiny BS-U} = \zeta_{u}^\text{\tiny BS-IRS} = \varrho d_{u}^{-\eta}$, where $d_{u}$ is the distance between the BS and the $u$th user and its serving IRS, $\varrho = 2\times 10^4$ is an array gain parameter \cite{SenaISIC2020}, and $\eta = 2$ is the path-loss exponent. Regarding the link IRS-U, since an IRS is a passive device, we model the corresponding fading coefficient as $\zeta_{gu}^\text{\tiny IRS-U} = \tilde{d}^{-\eta}$, where $\tilde{d} = 20$~m for all IRSs.

\begin{figure}[t]
	\includegraphics[scale=0.4]{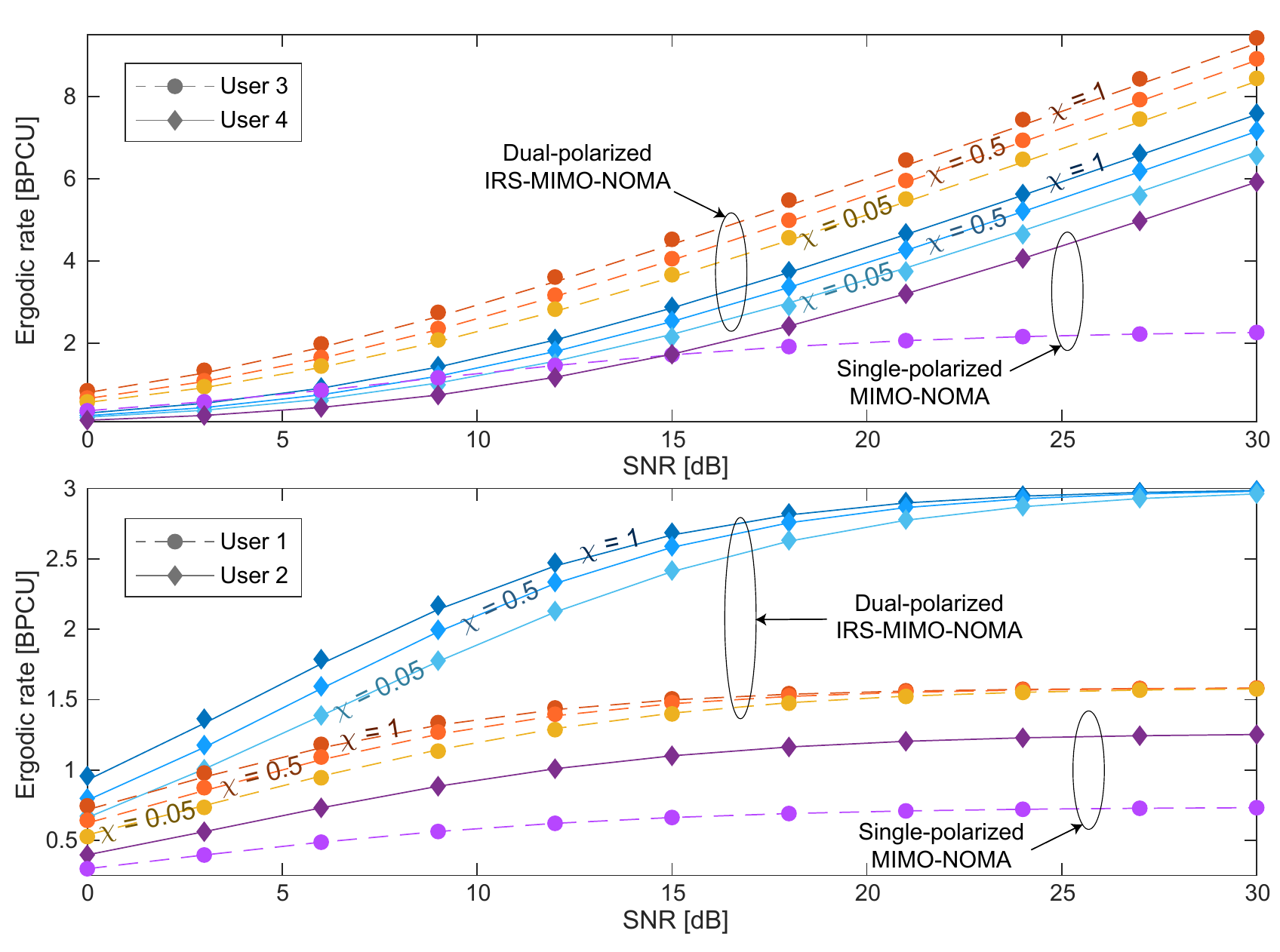}
	\centering
	\caption{Ergodic rates for different values of $\chi$ ($L = 500, N = 4, \xi= 0$).}\label{f3}
\end{figure}

By considering a large number of reflecting elements and perfect SIC, Fig. \ref{f3} presents the users' ergodic rates, generated by $R_u = E[\log_2(1 + \gamma_{u}^{u})]$. The figure confirms that, with the help of IRSs, depolarization phenomena can be transformed into an advantage, e.g., the higher the iXPD, the greater the performance gains. For instance, when user $3$ is served via the IRS-MIMO-NOMA scheme, for a low iXPD of $\chi = 0.05$, and an SNR of $30$~dB, its rate can reach $8.44$~BPCU, which is more than three times greater than that achieved in the single-polarized scheme. When we consider a high iXPD of $\chi = 1$, the achievable ergodic rate of the user $3$ becomes even more remarkable, reaching up to $9.42$~BPCU. Impressive performance gains can be also observed in all the other users.

\begin{figure}[t]
	\includegraphics[scale=0.4]{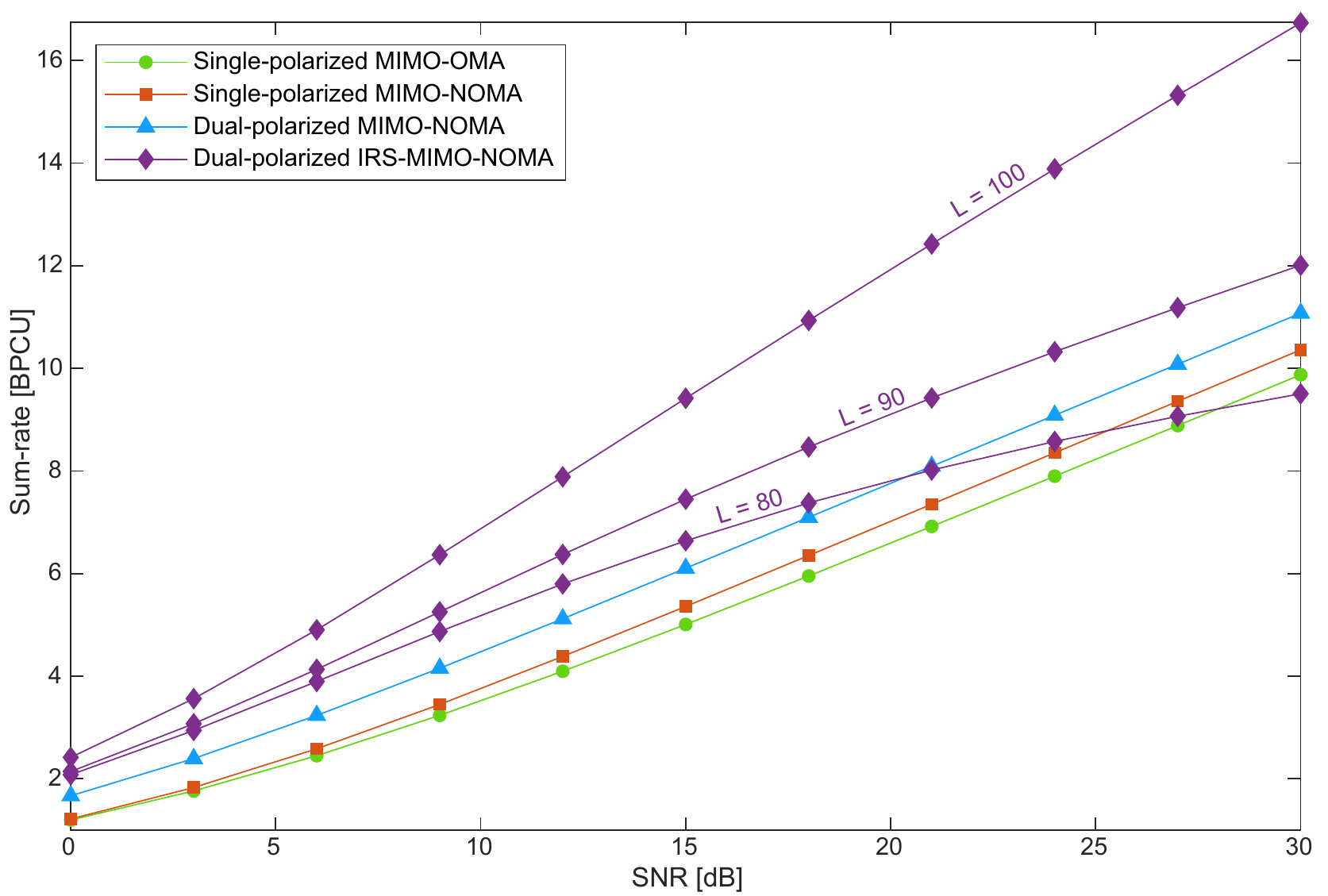}
	\centering
	\caption{Ergodic sum-rates. Comparison between the dual-polarized IRS-MIMO-NOMA and conventional schemes ($N = 4, \chi = 0.5, \xi= 0$).\vspace{-2mm}}\label{f5}
\end{figure}

\begin{figure}[t]
	\includegraphics[scale=0.4]{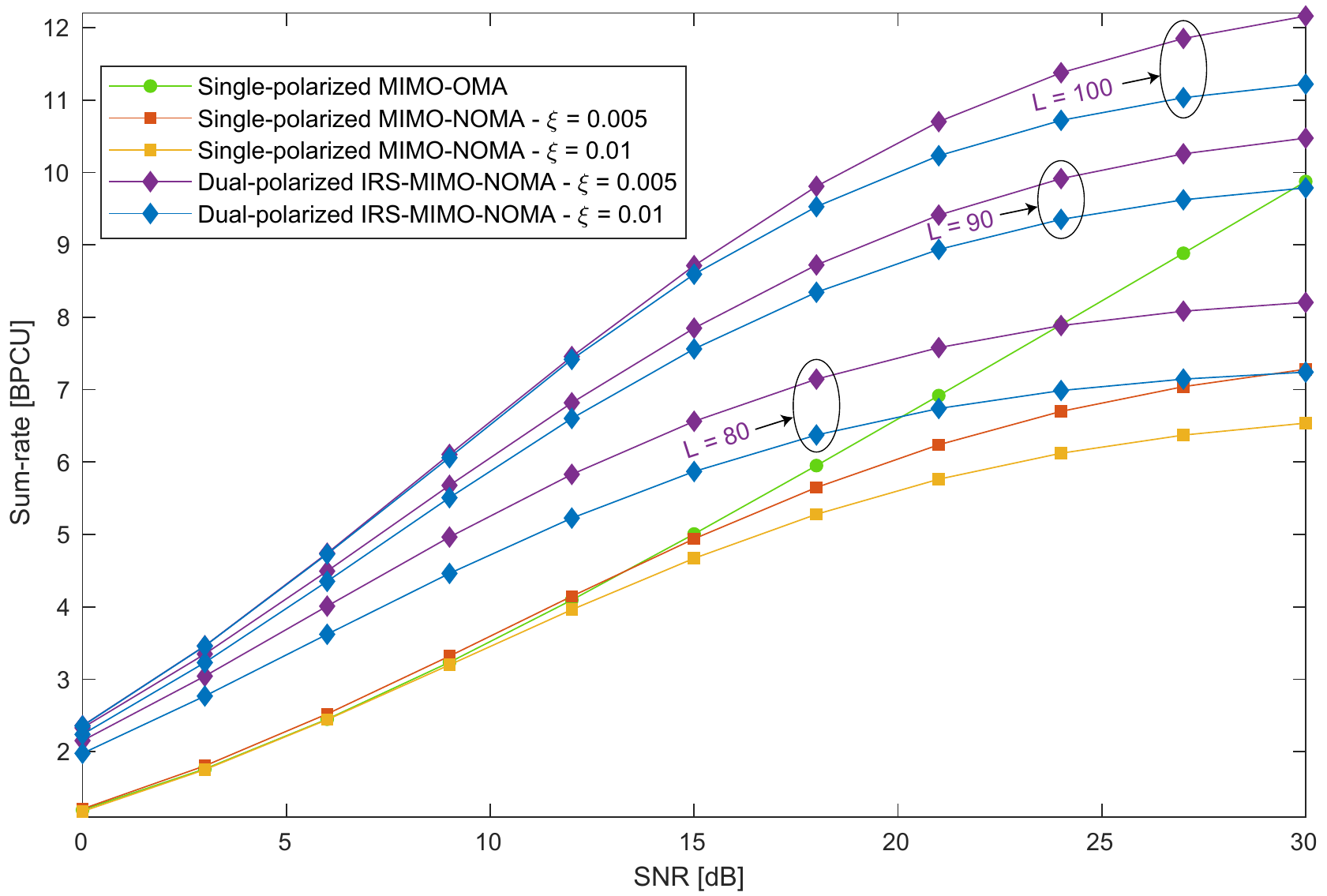}
	\centering
	\caption{Ergodic sum-rates. Comparison between the dual-polarized IRS-MIMO-NOMA and conventional schemes under imperfect SIC ($N = 4, \chi = 0.5$).}\label{f6}
\end{figure}

In Fig. \ref{f5}, we compare the sum-rates of the proposed IRS-MIMO-NOMA scheme and conventional systems assuming perfect SIC. As one can see, when $L=80$, from $21$~dB onward, the proposed scheme is outperformed by the dual-polarized MIMO-NOMA counterpart, and when the SNR reaches $30$~dB, the MIMO-OMA system is the one that achieves the best performance. However, with a slight increase in the number of reflecting elements, from $L = 80$ to $L = 90$, the IRS-MIMO-NOMA scheme outperforms all the other baseline schemes, in all considered SNR range. Finally, Fig. \ref{f6} shows how well the dual-polarized IRS-MIMO-NOMA system performs in comparison with the single-polarized MIMO-OMA and MIMO-NOMA counterparts in the presence of SIC error propagation. As can be seen, even though the sum-rate of all NOMA-based schemes are caped in the high-SNR regime, the proposed IRS-MIMO-NOMA system is significantly more robust to SIC errors than the conventional single-polarized MIMO-NOMA. For instance, even when considering $L=80$ reflecting elements, and an error of $\xi = 0.01$, the IRS-MIMO-NOMA can reach sum-rates remarkably higher than those achieved by the conventional schemes, being outperformed by the MIMO-OMA scheme only in SNR values above $20$~dB, and when $L=90$ and $L=100$, the IRS-MIMO-NOMA scheme always achieves the best performance.

\section{Conclusions}
By exploiting dual-polarized IRSs, we proposed a novel strategy for improving the performance of dual-polarized massive MIMO-NOMA networks with imperfect SIC. The detailed design of precoding and reception matrices was provided, and an efficient procedure for optimizing the IRS reflecting elements was developed. Our numerical results revealed that the proposed dual-polarized IRS-MIMO-NOMA scheme can achieve remarkable performance gains over conventional systems, enabling users to enjoy polarization diversity.
\vspace{-2mm}

\appendices

\section{Proof of Lemma I}\label{ap1}
\renewcommand{\theequation}{A-\arabic{equation}}
\setcounter{equation}{0}

After analyzing \eqref{sigbsic4}, we can express the effective channel gain obtained at the users by $\mathcal{\ddot{h}}_{gu} = \max\{\mathcal{h}^{v}_{gu}, \mathcal{h}^{h}_{gu}\}=\max \{1/[\mathbf{H}^{\dagger v}_{gu}(\mathbf{H}^{\dagger v}_{gu})^H]_{gg}, 1/[\mathbf{H}^{\dagger h}_{gu}(\mathbf{H}^{\dagger h}_{gu})^H]_{gg} \}$, denote the polarization interference by $\mathfrak{X}_{gu} = \left|[\mathbf{H}^{\dagger \ddot{p}}_{gu}\mathbf{\underline{H}}_{gu}^{t \ddot{p}}\mathbf{x}^t]_{g} \right|^2$, and define $\rho = 1/\sigma^2_n$ as the SNR. With these definitions, the $u$th user from subset $\mathcal{U}^{p}$, $p\in \{v,h\}$, decodes the message to the $i$th user, $\min\{\mathcal{U}^{p}\} < i<u$, $i \in \mathcal{U}^{p}$, with the following SINR

\begin{align}\label{eqB3}
\gamma_{gu}^{i}\text{\footnotesize $ = \rho\mathcal{\ddot{h}}_{gu} \alpha_{gi}^2 
\Bigg[
\rho\mathcal{\ddot{h}}_{gu} \Bigg( \sum_{m=i+1}^{\max\{\mathcal{U}^{p}\}} \hspace{-2mm} \alpha_{gm}^2 
+ \xi \hspace{-4mm}\sum_{n=\min\{\mathcal{U}^{p}\}}^{i-1} \hspace{-4mm} \alpha_{gn}^2 + \mathfrak{X}_{gu} \Bigg) + 
1\Bigg]^{-1}$}.
\end{align}

Note that when the weakest user, i.e., the user corresponding to $\min\{\mathcal{U}^{p}\}$, detects its symbol, it will experience interference from everyone else, but not errors from imperfect SIC. On the other hand, when the user with the best channel gain, i.e., the maximum index in $\mathcal{U}^{p}$, decodes its symbol, there will be no interference from higher-order users, but only from SIC errors. Therefore, the total SIC interference can be determined by 
\begin{align}
    \text{ $\mathfrak{I}_{gi} \hspace{-1mm}=$}\hspace{-1mm}
    \begin{cases}
        \sum_{m=i+1}^{\max\{\mathcal{U}^{p}\}} \alpha_{gm}^2, \hspace{-2mm} \quad \text{ if } i = \min\{\mathcal{U}^{p}\},\\
        \xi \sum_{n=1}^{i-1} \alpha_{gn}^2, \hspace{-2mm} \quad \text{ if } i = u = \max\{\mathcal{U}^{p}\},\\
        \hspace{-1mm} \sum_{m=i+1}^{\max\{\mathcal{U}^{p}\}} \hspace{-1mm} \alpha_{gm}^2 \hspace{-1mm} + \hspace{-.8mm}\xi \sum_{n=\min\{\mathcal{U}^{p}\}}^{i-1} \hspace{-1mm} \alpha_{gn}^2, \hspace{-2mm} \hspace{1mm} \text{ otherwise.}
    \end{cases}\nonumber
\end{align}

By applying the above definitions in \eqref{eqB3}, we can achieve the final SINR expression, which completes the proof. \hfill\qedsymbol

 \vspace{-.1cm}
\section*{Acknowledgments}
This work is partly supported by Academy of Finland via: (a) ee-IoT project under Grant n.319009, (b) FIREMAN consortium under Grant CHIST-ERA/n.326270, and (c) EnergyNet research fellowship under Grants n.321265 and n.328869. The work is also supported, in part, by EU H2020 RISE-6G project.

\ifCLASSOPTIONcaptionsoff
\newpage
\fi


\bibliographystyle{IEEEtran}
\bibliography{bibtex/references}

\end{document}